\documentclass[useAMS]{mn2e}
\usepackage{graphicx}

\title[A distance scale of PNe based on MIR data]
{A distance scale of planetary nebulae based on
mid-infrared data}
\author[R. Ortiz et al.]{R. Ortiz$^{1}$\thanks{E-mail:
ortiz@astro.iag.usp.br}, M.V.F. Copetti$^{2}$ and S. Lorenz-Martins$^{3}$ \\
$^{1}$Escola de Artes Ci\^encias e Humanidades, USP, Av. Arlindo Bettio, 1000,
S\~ao Paulo, SP, 03828-000, Brazil \\
$^{2}$Laborat\'orio de An\'alise Num\'erica e Astrof\'\i sica, Departamento
de Matem\'atica, UFSM, Santa Maria, RS, 97119-900, Brazil \\
$^{3}$Observat\'orio do Valongo, UFRJ, Ladeira do Pedro Ant\^onio 43,
Rio de Janeiro, RJ, 20080-090, Brazil}

\begin{document}

\date{Accepted. Received}


\maketitle

\label{firstpage}

\begin{abstract}

Some of the most successful statistical methods for obtaining distances
of planetary nebulae (PNe) are based on their apparent sizes and radio
emission intensities. These methods have the advantage of being
``extinction-free'' and are especially suited to be applied to PNe
situated at large distances. A similar method, based on the mid-infrared
(MIR) emission of PNe, would have the advantage of being applicable to
the large databases created after the various all-sky or Galactic plane
infrared surveys, such as IRAS, MSX, ISOGAL, GLIMPSE, etc. In this work
we propose a statistical method to calculate the distance of PNe based
on the apparent nebular radius and the MIR flux densities. We show that
the specific intensity between 8 and 21 micron is proportional to the
brightness temperature $T_b$ at 5 GHz. Using MIR flux densities at 8, 12,
15 and 21 microns from the MSX survey, we calibrate the distance
scale with a statistical method by Stanghellini
et al. 2008 (SSV). The database used in the calibration consisted of
67 Galactic PNe with MSX counterparts and distances determined by
SSV. We apply the method to a sample of PNe detected at 8 microns in the
GLIMPSE infrared survey, and determine the distance of a sample of PNe
located along the Galactic plane and bulge.

\end{abstract}


\section{Introduction}

The problem of determining the distances of Planetary Nebulae (PNe)
has long been
considered a cumbersome issue because of the numerous complexities
related to the evolution of these objects. Classical approaches to this
problem include trigonometric parallaxes (Harris et al. 1997),
the expansion method (Terzian 1997), spectroscopic distances
(M\'endez et al. 1988), etc. PNe nuclei are far from being
considered as ``standard candles'' since they evolve differently
according to their masses, a parameter that in most
cases is not easily determined. When the central star
is not visible, alternative methods based on properties of the nebula
itself are required. These methods are often
treated as ``statistical'', and are based on nebular properties
such as the ionized mass (Shklovsky 1956), the evolution of the nebular
optical depth with time (Kwok 1985, Marten \& Schoenberner 1991), the
electron density (Barlow 1987), etc.

The first statistical method proposed in the literature,
attributed to Shklovsky (1956), is based on the assumption
that the nebular ionized mass is an invariant. Although this hypothesis
has some limitations (Maciel \& Pottasch 1980), the method
represents a milestone to the distance problem, and was used
as a first step to calibrate other similar methods that followed (Daub 1982,
Cahn, Kaler \& Stanghellini 1992). Methods based on radio quantities
have several advantages over those derived from visual: the interstellar
extinction does not play a critical role, even in the cases where
the PNe are severely affected by interstellar extinction;
radio fluxes are available for a large number of
objects, etc. In addition to that, in the last three decades the increasing
number of observations carried out at high spatial resolution, such as
the VLA (Zijlstra et al. 1989), has largely improved our knowledge of young,
compact PNe. On the other hand, comparisons among some of these ``radio''
distance scales (Zhang 1995; Phillips 2002, 2004)
reveal that they often exhibit significant biases, showing that the
problem is still an open issue.

\begin{figure*}
\includegraphics[width=16.0cm]{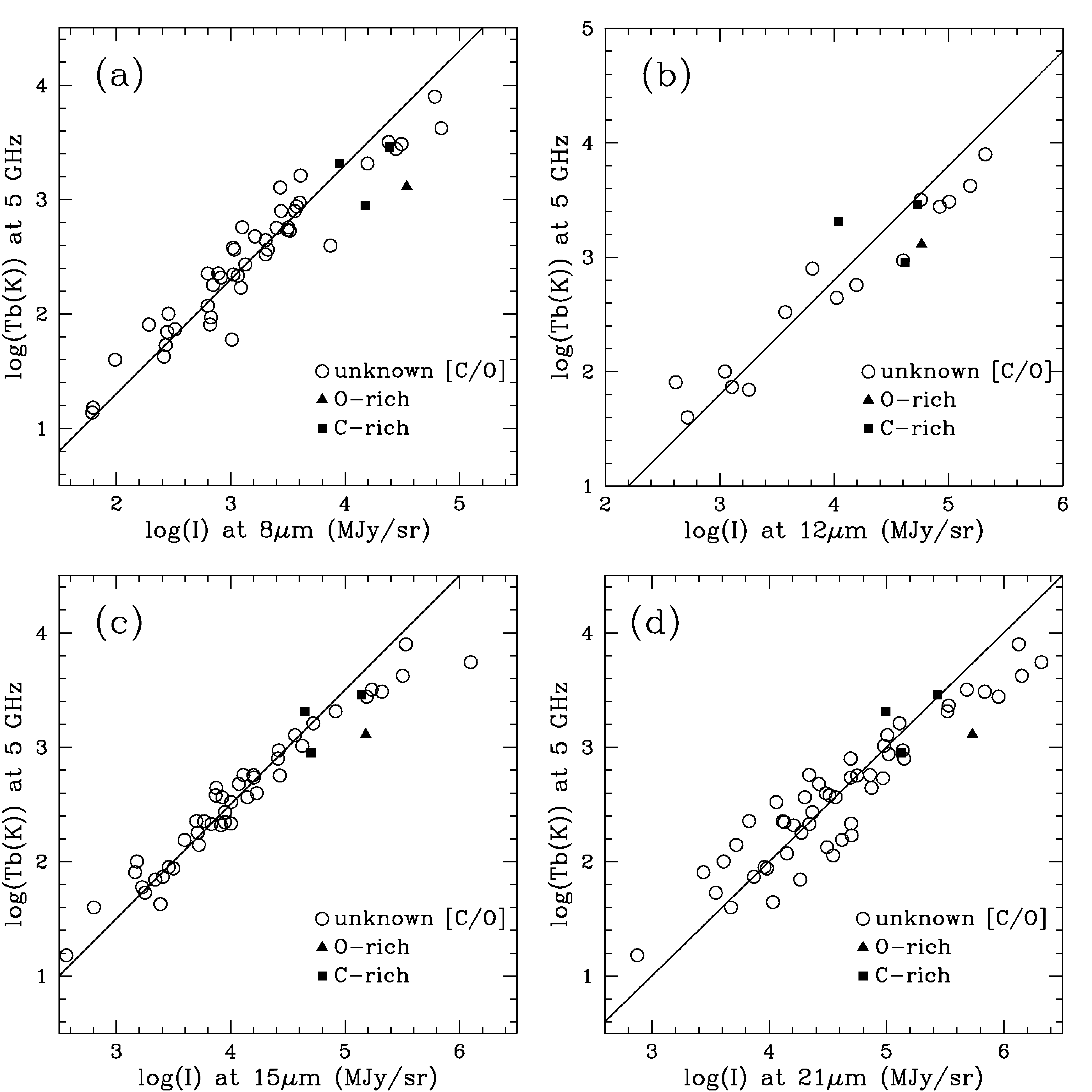}
\caption{Brightness temperature at 5 GHz as a function of the specific
intensity at 8 (a), 12 (b), 15 (c) and 21 (d) $\mu$m. The straight line
represents the best fit with slope
$\partial \log T_b /\partial \log I_{\lambda}=1$.
Open circles represent PNe without [C/O] data; filled triangles depict
O-rich PNe, classified according to their mid-infrared spectra; filled
squares represent PNe with carbonaceous features in their spectra.
PNe evolve from the top-right to the bottom-left corner of this figure:
the radio emission becomes optically thinner whilst the infrared intensity
decreases with time.}
\end{figure*}

Mid and far-infrared data constitute an interesting
alternative to approach the distance problem, because PNe are generally
strong infrared sources and their emission in this band is little affected
by the interstellar extinction. These two characteristics allow the detection
of PNe severely obscured by dust, such as PNe hidden in the galactic
plane, for example.
A relationship between the nebular emission at infrared and radio
wavelengths was established by Zhang \& Kwok (1993), who studied the
correlation between the {\sc IRAS} specific intensity at 60 $\mu$m ($I_{60}$)
and the brightness temperature ($T_{\rm b}$) at 5 GHz. These two
distance-independent parameters can be correlated with the nebular size
to compose a new distance scale. Unfortunately {\sc IRAS} fluxes suffer from
poor spatial resolution ($\sim 1$\arcmin) and the severe crowdness
at low Galactic latitudes, limitations that were partially overcame
by the {\sc MSX} survey (Egan et al. 1997, 1999), thanks to its better
spatial resolution ($\sim 18$\arcsec ). {\sc MSX} covered the Galactic
plane between $\vert b \vert < 5^o$ in 4 photometric bands, centred at
8, 12, 15 and 21 $\mu$m. Another milestone infrared survey was
{\sc ISOGAL} (Omont et al. 2003), a part of the ISO mission designed
to obtain mid-infrared images of the Galactic bulge and specific regions
of the Galactic plane with spatial resolution better than 6\arcsec\ mainly at
7 and 15 $\mu$m. Although its sensitivity was almost two orders of
magnitude deeper than {\sc IRAS} and {\sc MSX}, {\sc ISOGAL} suffered
from its small sky coverage, limited to only $\sim 16$ square degrees.
Nevertheless, the total number of detections, including point and
extended sources, amounts {\bf to} $\sim 10^5$ objects. More recently,
the Galactic plane and bulge were surveyed by GLIMPSE (Galactic Legacy
Infrared Mid-Plane Survey Extraordinaire, Fazio et al. 2004), a project
developed as a part of the ``Spitzer'' mission.
GLIMPSE was restricted to $\vert l \vert < 65^o$ and carried out
with the infrared camera ``IRAC'', which obtained images simultaneously
in four mid-infrared bands, centred at 3.6, 4.5, 5.8 and 8.0 $\mu$m
and with spatial resolution better than 2\arcsec . These surveys
contain hundreds of infrared counterparts of PNe that can be used
as an efficient tool to derive the distance of PNe severely affected
by interstellar extinction, especially those situated along the
Galactic plane.

\begin{table}
\centering
\caption{The {\sc MSX} bands and the coefficients of equation 4.}
\begin{tabular}{cccccc}
\hline

Band & ${\lambda}_{\mathrm iso}$($\mu$m) & a$_{\lambda}$ & b$_{\lambda}$ &
c$_{\lambda}$ & N \\

\hline

A &   8.28 & $-$0.1736 & $-$0.3899 & +0.7960 & 59 \\
C &  12.13 & $-$0.2728 & $-$0.2551 & +0.8527 & 15 \\
D &  14.65 & $-$0.2491 & $-$0.3291 & +0.9217 & 56 \\
E &  21.34 & $-$0.2382 & $-$0.3242 & +1.0342 & 52 \\

\hline
\end{tabular}
\end{table}

The main purpose of the present study is to derive a new distance scale
based on a minimal set of data: mid-infrared intensities and the
angular size. Although the present calibration
is based on {\sc MSX} flux densities (at 8, 12, 15 and 21 $\mu$m), it can be
extended to other existing data extracted from past (such as {\sc IRAS},
{\sc ISOGAL}, GLIMPSE) or future surveys carried out at similar wavelengths.

This paper is organized as follows: Sect. 2 contains a general discussion
of the radio and infrared emission of PNe and the definition of basic
concepts used in the following sections; Sect. 3 presents the formulation of
the method and its calibration; Sect. 4 is devoted to show the results,
including an application of the method to PNe detected by {\it Spitzer};
our conclusions are presented in Sect. 5.

\section{The radio and infrared emission of planetary nebulae}

Since the formation of a planetary nebula succeeds the Asymptotic
Giant Branch (AGB), much of the circumstellar dust formed during this
latter phase survives after the onset of the ionization of the
nebula. The first detection of infrared radiation of PN origin is
credited to Gillet et al. (1967) and since then, hundreds of
infrared counterparts of PNe have been detected.

The main characteristics of nebular spectra in the mid-infrared are:
{\it (i) the continuum}, that results from the thermal emission of
heated grains in the nebula; it appears stronger in newly formed PNe
and decreases in intensity as the nebula evolves, as shown in Fig. 1
{\it (ii) atomic emission lines}, emitted during transitions between fine
structure atomic levels; a few of them are strong enough to affect
broad-band magnitudes between 8 and 25 $\mu$m, the most prominent are:
[Ar\,{\small\sc III}] $9.0\,\mu$m, 
[S\,{\small\sc IV}] $10.53\,\mu$m, 
[Ne\,{\small\sc II}] $12.8\,\mu$m,
[Ne\,{\small\sc V}] $14.3\,\mu$m, 
[Ne\,{\small\sc III}] $15.4\,\mu$m, 
[S\,{\small\sc III}] $18.7\,\mu$m,
[Ar\,{\small\sc III}] $21.8\,\mu$m and 
[Ne\,{\small\sc V}] $24.1\,\mu$m; 
{\it (iii) molecular bands}, in emission or absorption, are generally
caused by vibration transitions of molecules present in ice mantles
deposited on the grains; they appear as broad emission features
such as the amorphous silicate band at 9.7 $\mu$m (due to Si$-$O
stretching), SiC at 11.2 $\mu$m, the numerous ice and PAH features
between 2 and 20 $\mu$m, etc.

Among the hypotheses adopted in the present method is that the
infrared specific intensity of a PN is related to
the evolutionary status of the nebula and does not depend on its
distance. This was formerly observed by Kwok (1989) and Zhang
\& Kwok (1990), who studied the {\sc IRAS} 60 $\mu$m emission of
PNe. The specific intensity can be evaluated by the following equation:

\begin{equation}
I_{\lambda}= 1.354 \times 10^4 \frac{F_{\lambda}}{{\theta}^2},
\end{equation}

\noindent where $I_{\lambda}$(MJy sr$^{-1}$) is the specific
intensity, $F_{\lambda}$ is the infrared flux density (in Jansky),
and $\theta$ is the nebular radius, in arcseconds. The specific
intensity given by the equation above represents only an average
value across the PNe, which in all cases show variations across
the surface. Another hypothesis assumed is the spherical symmetry,
even though it can be valid also for elliptical nebulae since ${\theta}^2$
represents to the product of the nebular major and minor semi-axes.

This work is based on the the mid-infrared flux densities
provided by the {\sc MSX} Survey. This choice is justified by
the following facts: {\it(i)} the large number of {\sc MSX}
counterparts of PNe and the homogeneousness of those data; {\it (ii)}
the better spatial resolution of the {\sc MSX} survey, compared to
{\sc IRAS}; {\it (iii)} the variety of
photometric bands (4) between 8 and 21 $\mu$m. The database adopted in
this study is the list of {\sc MSX} counterparts of PNe given by Ortiz
et al. (2005). The list (table 3 of that paper) contains 216 objects,
all of them smaller than the spatial resolution of the ``SPIRIT'' camera
of 18.3\arcsec . Therefore, all the nebulae included in this study
are compact from the {\sc MSX} point of view, and their photometry
includes the flux density of the whole nebula.

The radio emission of a PN is generally expressed in terms of
{\it brightness temperature}, which is proportional to the radio
intensity at 5 GHz, as follows:

\begin{equation}
T_b = 17.67 \frac{S_{\rm 5 GHz}}{{\theta}^2},
\end{equation}

\noindent where $T_b$ is given in Kelvin, $S_{\rm 5 GHz}$ in miliJansky
(mJy), and the apparent radius $\theta$ in arc seconds. Also here
the brightness temperature represents only an average value across the
nebula, and the same considerations made for equation 1 must be
observed.

In this work, all $T_b$ data have been extracted from the paper
of Si\'odmiak \& Tylenda (2001), which represents a compilation
of 264 objects observed
at 1.4 and 5 GHz, all of them using the VLA. Thus, the full set
of radio data, including $T_b$ and $\theta$, have been obtained
with the same set of radiotelescopes, a procedure that helps
to assure data homogeneity.

Before we begin to devise the distance scale, we shall examine a
relationship between these two distance-independent parameters:
$T_b$ at 5 GHz and the infrared nebular specific intensity $I_{\lambda}$.
According to Zhang \& Kwok (1993), who studied {\sc IRAS} counterparts of
PNe observed with the VLA, these two quantities are well correlated with
each other.
In Fig. 1, the brightness temperature $T_b$ at 5 GHz is plotted against
the {\sc MSX} specific intensities at 8, 12, 15 and 21 $\mu$m. The data
follow a descending track from the top-right to the bottom-left corner,
as firstly shown by Volk (1992) and Zhang \& Kwok (1993), who used
theoretical evolutionary models of the central star (Sch\"onberner 1981,
1983; Bl\"ocker \& Sch\"onberner 1990) to perform radiative transfer
calculations of the parameters of the post-AGB envelope and the
newly-formed nebula. The brightness temperature $T_b$ decreases with
time because of the decreasing
optical depth of the radiation at 5 GHz as the nebula expands. Therefore,
nebulae situated near the top-right corner are younger and optically
thick at 5 GHz, whereas those near the bottom-left corner are older
and their continuum radiation emitted at 5 GHz is optically thin.
One can see that the maximum $T_b$ value (the optically thick case)
in our sample is about $1.0 \times 10^4$K, which corresponds approximately
to the typical electron temperature of a PN. The sequence ends near
$T_b \simeq 15$ K, whereas in Zhang \& Kwok (1993) it goes down to
$T_b \simeq 2 \times 10^{-2}$ K. This important difference is caused by
the fact that the PNe selected for the present study, taken from Ortiz
et al. (2005), are compact and restricted to objects with angular
diameter below the {\sc MSX} spatial resolution of 18.3\arcsec. This constraint
introduces a bias towards younger objects, which generally exhibit larger
$T_b$ values. As for the central star, little can be inferred from its
position in the diagram: Zhang \& Kwok (1993) showed that the evolutionary
track followed by central stars of different masses is practically the
same, even though there are significant differences among the speed that
central stars of different masses cross the diagram. The correlation
between $I_\lambda$ and $T_b$ shown in Fig. 1 confirms that the {\sc MSX}
intensity is proportional to the brightness temperature at 5 GHz over
a wide range of optical depths and this is expected to be valid
for other surveys carried out at similar wavelenghts.

\begin{table*}
\centering
\caption{Distances of the calibrators of Stanghellini et al. (2008, SSV)
that appear also in the list of Ortiz et al. (2005), in alphabetical order.}
\begin{tabular}{clcrrrrrcccccr}
\hline
\#  & PN & {\sc MSX} & $\theta$ & F$_8$ & F$_{12}$ &
F$_{15}$ & F$_{21}$ & D$_{8}$ & D$_{12}$ & D$_{15}$ &
D$_{21}$ & D$_{\rm ave}$ & D$_{\rm SSV}$ \\
 & name & name & (\arcsec ) & (Jy) & (Jy) & (Jy) & (Jy) &
(kpc) & (kpc) & (kpc) & (kpc) & (kpc) & (kpc) \\

\hline

 22 & Ap1-12       & 003.3253$-$04.6590 &  6.00 &   0.15 &    -   &    -   &   5.75 &  4.34 &   -   &   -   &  3.99 &  4.17 &  4.61 \\
 20 & Hb4          & 003.1733+02.9275 &  3.63 &   0.76 &    -   &   4.88 &   6.58 &  3.97 &   -   &  3.68 &  4.55 &  4.07 &  5.04 \\
178 & Hb5          & 359.3566$-$00.9801 &  1.70 &   5.22 &  11.43 &  29.73 &  58.04 &  3.82 &  3.20 &  3.01 &  3.46 &  3.37 &  1.69 \\
 82 & He2-11       & 259.1519+00.9404 & 32.50 &   0.24 &    -   &   1.27 &    -   &  2.06 &   -   &  2.50 &   -   &  2.28 &  0.89 \\
 84 & He2-15       & 261.6163+03.0039 & 11.90 &   0.13 &    -   &    -   &    -   &  3.39 &   -   &   -   &   -   &  3.39 &  2.17 \\
 98 & He2-85       & 300.5892$-$01.1090 &  5.10 &   0.44 &    -   &   2.02 &   5.24 &  3.81 &   -   &  4.10 &  4.30 &  4.07 &  3.56 \\
102 & He2-99       & 309.0023$-$04.2404 &  8.50 &   0.23 &    -   &   2.39 &   6.48 &  3.49 &   -   &  3.32 &  3.46 &  3.43 &  3.77 \\
106 & He2-111      & 315.0311$-$00.3707 &  6.00 &   0.20 &    -   &   1.48 &    -   &  4.10 &   -   &  4.20 &   -   &  4.15 &  3.51 \\
112 & He2-123      & 323.9552+02.4585 &  2.30 &   0.35 &    -   &   1.05 &   3.46 &  5.42 &   -   &  6.28 &  6.14 &  5.95 &  5.75 \\
114 & He2-125      & 324.2651+02.5883 &  1.50 &   0.12 &    -   &   0.95 &   3.46 &  7.76 &   -   &  7.40 &  7.06 &  7.41 & 10.41 \\
111 & He2-132      & 323.1214$-$02.5639 &  8.90 &   0.12 &    -   &   1.36 &   2.86 &  3.86 &   -   &  3.77 &  4.15 &  3.92 &  3.43 \\
116 & He2-141      & 325.4953$-$04.0062 &  6.90 &   0.13 &    -   &   1.30 &    -   &  4.19 &   -   &  4.14 &   -   &  4.17 &  3.47 \\
118 & He2-142      & 327.1953$-$02.2045 &  1.80 &   3.96 &  10.06 &  10.63 &  23.61 &  3.92 &  3.27 &  3.82 &  4.21 &  3.80 &  7.40 \\
119 & He2-143      & 327.8487$-$01.6702 &  2.60 &   0.74 &   1.44 &   4.82 &  13.71 &  4.54 &  5.06 &  4.12 &  4.25 &  4.49 &  5.25 \\
128 & He2-152      & 333.4300+01.1870 &  5.50 &   1.13 &    -   &   6.64 &  11.00 &  3.15 &   -   &  2.97 &  3.52 &  3.21 &  3.04 \\
124 & He2-161      & 331.5436$-$02.7880 &  5.00 &   0.12 &    -   &   0.93 &    -   &  4.83 &   -   &  5.00 &   -   &  4.92 &  5.31 \\
126 & He2-164      & 332.0796$-$03.3613 &  8.00 &   0.12 &    -   &    -   &   3.21 &  3.99 &   -   &   -   &  4.18 &  4.08 &  2.79 \\
138 & IC4637       & 345.4789+00.1411 &  9.30 &    -   &    -   &    -   &  10.91 &   -   &   -   &   -   &  2.97 &  2.97 &  2.40 \\
 23 & IC4673       & 003.5529$-$02.4423 &  8.50 &   0.33 &    -   &   1.96 &   4.01 &  3.28 &   -   &  3.49 &  3.88 &  3.55 &  3.22 \\
 71 & K3-7         & 028.7768+02.7024 &  3.15 &   0.20 &    -   &   1.31 &   2.58 &  5.29 &   -   &  5.35 &  5.95 &  5.53 &  6.12 \\
 65 & K3-11        & 023.8099$-$01.7905 &  1.50 &    -   &    -   &    -   &   5.18 &   -   &   -   &   -   &  6.41 &  6.41 & 10.80 \\
 74 & K3-18        & 032.0159$-$03.0364 &  2.00 &   2.28 &   3.75 &   4.64 &  14.91 &  4.14 &  4.16 &  4.53 &  4.54 &  4.34 &  9.60 \\
 30 & M1-25        & 004.9383+04.9365 &  1.60 &   0.20 &    -   &   1.40 &   6.17 &  6.89 &   -   &  6.58 &  6.02 &  6.50 &  6.61 \\
158 & M1-27        & 356.5305$-$02.3940 &  4.00 &   0.33 &   2.13 &   2.60 &  21.75 &  4.42 &  4.07 &  4.17 &  3.31 &  3.99 &  4.61 \\
175 & M1-29        & 359.1141$-$01.7195 &  3.50 &    -   &    -   &   4.78 &   4.75 &   -   &   -   &  3.75 &  4.97 &  4.36 &  4.13 \\
 38 & M1-31        & 006.4554+02.0150 &  3.50 &    -   &    -   &   2.86 &   8.68 &   -   &   -   &  4.26 &  4.31 &  4.28 &  5.10 \\
 27 & M1-35        & 003.9229$-$02.3230 &  2.20 &   0.38 &    -   &   3.18 &   4.82 &  5.44 &   -   &  4.83 &  5.76 &  5.35 &  6.09 \\
 53 & M1-39        & 015.9275+03.3589 &  2.00 &   0.60 &   3.12 &   2.21 &  22.05 &  5.22 &  4.38 &  5.46 &  4.14 &  4.80 &  6.37 \\
 40 & M1-41        & 006.7679$-$02.2531 & 38.00 &   0.50 &    -   &   3.02 &   3.67 &  1.71 &   -   &  1.92 &  2.44 &  2.02 &  0.85 \\
 55 & M1-46        & 016.4510$-$01.9754 &  6.00 &   0.26 &   1.39 &   1.70 &  12.57 &  3.93 &  4.12 &  4.06 &  3.31 &  3.86 &  3.58 \\
 52 & M1-50        & 014.6219$-$04.3797 &  2.80 &   0.19 &    -   &   1.19 &   3.62 &  5.59 &   -   &  5.70 &  5.70 &  5.66 &  5.98 \\
 61 & M1-51        & 020.9993$-$01.1252 &  7.50 &   1.19 &   4.61 &   6.30 &  17.03 &  2.77 &  2.81 &  2.72 &  2.87 &  2.79 &  2.29 \\
 54 & M1-54        & 016.0655$-$04.3900 &  6.50 &   0.13 &    -   &   1.41 &    -   &  4.29 &   -   &  4.14 &   -   &  4.22 &  3.81 \\
 63 & M1-57        & 022.1811$-$02.4085 &  4.00 &   0.39 &   1.51 &   3.02 &   8.76 &  4.29 &  4.47 &  4.02 &  4.12 &  4.22 &  4.40 \\
 62 & M1-58        & 022.0690$-$03.1846 &  3.20 &   0.24 &    -   &   1.15 &    -   &  5.09 &   -   &  5.49 &   -   &  5.29 &  5.32 \\
 67 & M1-59        & 023.9280$-$02.3440 &  2.40 &   0.86 &   1.59 &   4.30 &   4.89 &  4.56 &  5.02 &  4.35 &  5.58 &  4.88 &  5.42 \\
 59 & M1-60        & 019.7946$-$04.5277 &  1.25 &   0.36 &    -   &   1.84 &   5.74 &  6.84 &   -   &  6.66 &  6.64 &  6.71 &  9.41 \\
 75 & M1-66        & 032.7856$-$02.0419 &  1.35 &   0.17 &    -   &   1.74 &   2.95 &  7.57 &   -   &  6.59 &  7.58 &  7.25 &  8.77 \\
160 & M2-11        & 356.9514+04.5575 &  1.35 &    -   &    -   &   0.91 &   2.97 &   -   &   -   &  7.74 &  7.57 &  7.66 & 10.75 \\
162 & M2-16        & 357.4824$-$03.2479 &  1.10 &   0.19 &    -   &   1.25 &   3.30 &  8.05 &   -   &  7.65 &  7.89 &  7.86 &  7.04 \\
163 & M2-18        & 357.4094$-$03.5480 &  0.75 &   0.14 &    -   &    -   &   3.88 &  9.88 &   -   &   -   &  8.60 &  9.24 & 15.74 \\
  6 & M2-21        & 000.7012$-$02.7613 &  1.50 &   0.18 &    -   &    -   &    -   &  7.18 &   -   &   -   &   -   &  7.18 & 10.16 \\
 14 & M2-23        & 002.2269$-$02.7854 &  4.40 &   0.71 &   1.51 &   3.13 &   8.98 &  3.72 &  4.37 &  3.86 &  3.97 &  3.98 &  4.26 \\
 26 & M2-30        & 003.7894$-$04.6562 &  1.75 &   0.15 &    -   &    -   &    -   &  6.99 &   -   &   -   &   -   &  6.99 &  9.44 \\
 70 & M2-44        & 028.5976+01.6550 &  4.00 &   1.21 &    -   &   1.99 &    -   &  3.52 &   -   &  4.46 &   -   &  3.99 &  4.76 \\
 69 & M2-45        & 027.7030+00.7052 &  3.20 &   0.69 &    -   &   4.96 &   8.66 &  4.24 &   -   &  3.82 &  4.44 &  4.17 &  4.50 \\
161 & M3-7         & 357.1186+03.6107 &  2.35 &    -   &    -   &   1.18 &   3.72 &   -   &   -   &  6.05 &  6.00 &  6.03 &  6.32 \\
166 & M3-10        & 358.2412+03.6343 &  1.50 &   0.18 &    -   &   1.40 &   3.35 &  7.20 &   -   &  6.72 &  7.11 &  7.01 &  9.33 \\
152 & M3-14        & 355.4412$-$02.4679 &  1.40 &   0.19 &    -   &   1.29 &   3.39 &  7.29 &   -   &  7.01 &  7.25 &  7.18 &  5.70 \\
  8 & M3-23        & 000.9220$-$04.8521 &  6.00 &   0.16 &    -   &    -   &    -   &  4.25 &   -   &   -   &   -   &  4.25 &  4.25 \\
159 & M3-38        & 356.9847+04.4410 &  0.90 &   0.24 &    -   &   1.47 &   4.41 &  8.33 &   -   &  7.86 &  7.86 &  8.01 & 14.06 \\
  2 & M3-43        & 000.1231$-$01.1453 &  1.90 &   0.20 &    -   &   1.54 &   4.25 &  6.41 &   -   &  6.07 &  6.22 &  6.24 &  8.53 \\
 90 & My60         & 283.8976+02.2686 &  3.80 &   0.18 &    -   &   1.01 &   2.40 &  4.99 &   -   &  5.36 &  5.70 &  5.35 &  4.80 \\
110 & Mz1          & 322.4909$-$02.6123 & 12.90 &   0.20 &    -   &   1.36 &    -   &  3.05 &   -   &  3.34 &   -   &  3.20 &  2.30 \\
121 & Mz2          & 329.3843$-$02.8720 & 11.50 &   0.16 &    -   &   1.10 &    -   &  3.32 &   -   &  3.65 &   -   &  3.48 &  2.36 \\
125 & Mz3          & 331.7287$-$01.0105 & 12.70 &  38.79 &  80.40 &  93.19 & 311.80 &  1.23 &  1.13 &  1.17 &  1.21 &  1.18 &  1.45 \\
 77 & NGC2440      & 234.8379+02.4206 &  9.00 &   1.16 &   2.47 &   8.75 &  16.49 &  2.59 &  3.18 &  2.36 &  2.72 &  2.71 &  1.36 \\
 79 & NGC2452      & 243.3792$-$01.0384 &  9.40 &   0.22 &    -   &    -   &   3.29 &  3.39 &   -   &   -   &  3.94 &  3.67 &  2.84 \\
 86 & NGC2792      & 265.7520+04.1025 &  6.50 &   0.27 &    -   &   1.50 &   4.77 &  3.79 &   -   &  4.08 &  4.06 &  3.98 &  3.05 \\
141 & NGC6302      & 349.5082+01.0551 & 22.30 &  13.60 &  23.77 &  71.78 & 261.30 &  1.18 &  1.36 &  1.04 &  1.05 &  1.16 &  0.74 \\

\hline
\end{tabular}
\end{table*}

\setcounter{table}{1}
\begin{table*}
\centering
\caption{continued}
\begin{tabular}{clcrrrrrcccccr}
\hline
\#  & PN & {\sc MSX} & $\theta$ & F$_8$ & F$_{12}$ &
F$_{15}$ & F$_{21}$ & D$_{8}$ & D$_{12}$ & D$_{15}$ &
D$_{21}$ & D$_{\rm ave}$ & D$_{\rm SSV}$ \\
 & name & name & (\arcsec ) & (Jy) & (Jy) & (Jy) & (Jy) &
(kpc) & (kpc) & (kpc) & (kpc) & (kpc) & (kpc) \\

\hline

 43 & NGC6445      & 008.0752+03.9075 & 16.60 &   0.29 &    -   &   2.95 &   4.95 &  2.60 &   -   &  2.53 &  2.97 &  2.70 &  1.38 \\
 24 & NGC6565      & 003.5326$-$04.6223 &  4.50 &   0.19 &    -   &   1.16 &   3.35 &  4.64 &   -   &  4.90 &  4.98 &  4.84 &  4.66 \\
 49 & NGC6567      & 011.7433$-$00.6498 &  4.40 &   1.54 &   2.65 &   4.00 &   5.34 &  3.26 &  3.74 &  3.63 &  4.49 &  3.78 &  3.61 \\
 47 & NGC6578      & 010.8183$-$01.8272 &  4.30 &    -   &    -   &   7.15 &  13.28 &   -   &   -   &  3.17 &  3.64 &  3.40 &  3.64 \\
 87 & PB3          & 269.7392$-$03.6199 &  3.50 &   0.18 &    -   &   1.31 &   2.81 &  5.18 &   -   &  5.17 &  5.63 &  5.33 &  4.89 \\
129 & Pe1-6        & 336.2217+01.9726 &  3.60 &    -   &    -   &   1.07 &    -   &   -   &   -   &  5.39 &   -   &  5.39 &  5.38 \\
 19 & Pe2-12       & 002.8530$-$02.2915 &  2.50 &   0.22 &    -   &    -   &    -   &  5.69 &   -   &   -   &   -   &  5.69 & 12.19 \\

\hline
\end{tabular}
\end{table*}

\subsection{Mid-infrared spectra of PNe: the role of the [C/O] ratio}

The mid-infrared spectra of PNe show large differences from object
to object and the characteristics of a PN are very difficult to
predict because of the large number of factors that play a
role in the composition of the spectrum. For example,
in a study of $\sim 500$ PNe carried out by Stasi\'nska \& Szcerba
(1999) based on {\sc IRAS} data, a large range of dust-to-gas mass ratios
was observed, even though this parameter seemed not to vary as the
nebula evolves. The same study also showed that the contamination
of the {\sc IRAS} broad-band fluxes by atomic lines
is more evident in the [12] and [25] bands, especially among the
more evolved nebulae. Other papers confirm the difficulty in
disentangling the individual contributions of stellar, nebular and
dust contribution to the total nebular spectrum (Volk 1992,
Zhang \& Kwok 1991). Most of these studies are based on models
intended to generate synthetic spectra according to a grid of
various parameters such as the temperature of the central star,
electron density, dust-to-gas ratio, chemical abundances, etc.

The effect of the chemical abundances on the final infrared spectrum
is also difficult to evaluate because they do not only affect the
intensities of atomic lines but also change the balance of
energy in the nebula. Oxygen lines for instance act as a cooler of the
nebula that eventually decrease the thermal emission of dust.
Understanding the role of the abundances in the
nebular spectrum is also difficult because the abundances
of certain elements, such as carbon for example, are very difficult
to determine because of the lack of atomic lines in the visible region
of the spectrum. In addition to that, a few chemical elements show strong
atomic lines in the mid-infrared, the most important are S, Ne and Ar.

In a study of 25 PNe seen against the Galactic bulge, Casassus et al.
(2001a) classified PNe according to the [C/O] ratio using molecular features
identified in mid-infrared spectroscopy between $8-13\,\mu$m.
Among the 13 bulge members with identified dust types, 30\% were associated
with carbonaceous grains, whilst the remaining 70\% showed SiO features
normally associated with O-rich grains. However, the same methodology
applied to 54 PNe distributed throughout the Galactic disc resulted
that 30\% of all chemically identified PNe inside the Solar circle
are O-rich PNe, but this percentage is only 14\% outside it (Casassus
et al. 2001b). Therefore, the relative [C/O] nebular abundance shows
a radial gradient similar to what is observed in the general distribution
of carbon and oxygen-rich AGB stars (OH/IR, for example, Blommaert
et al. 1993, Guglielmo et al. 1998).

In this study we attempt to use the [C/O] mid-infrared classification to
verify if the [C/O] ratio produces significant deviations from the general
evolution of PNe, with respect to their broad-band mid-infrared fluxes.
Unfortunately, the number of PNe with carbonaceous or silicate
features identified in their mid-infrared spectra is very small,
and our sample contains only 20 objects classified according to
this criterium. Among them, only 4 nebulae have their sizes accurately
determined by the VLA: NGC6537, M1-12, Hb-5 (all C-rich), and H1-40
(O-rich). Their position at the upper part of the evolutionary track
seen in the $I_{\lambda}$ versus $T_b$ diagram (Fig. 1) indicates that
they are among the youngest objects in the sample. Actually, the fact that
they all show a strong continuum is a direct consequence of their
early evolutionary status. As the PN nucleus (PNN) evolves, the
circumstellar dust is slowly destroyed by the UV photons originated
from the PNN. Therefore, since the observation of molecular features
in the mid-infrared requires the presence of a strong continuum, it
is straightforward to conclude that any classification based on
these features is biased towards the less evolved PNe.
Because of the small number of objects with mid-infrared spectra,
little can be concluded about the role of the [C/O] ratio in the
$I_{\lambda}$ versus $T_b$ relationship.

\begin{figure*}
\includegraphics[width=16.0cm]{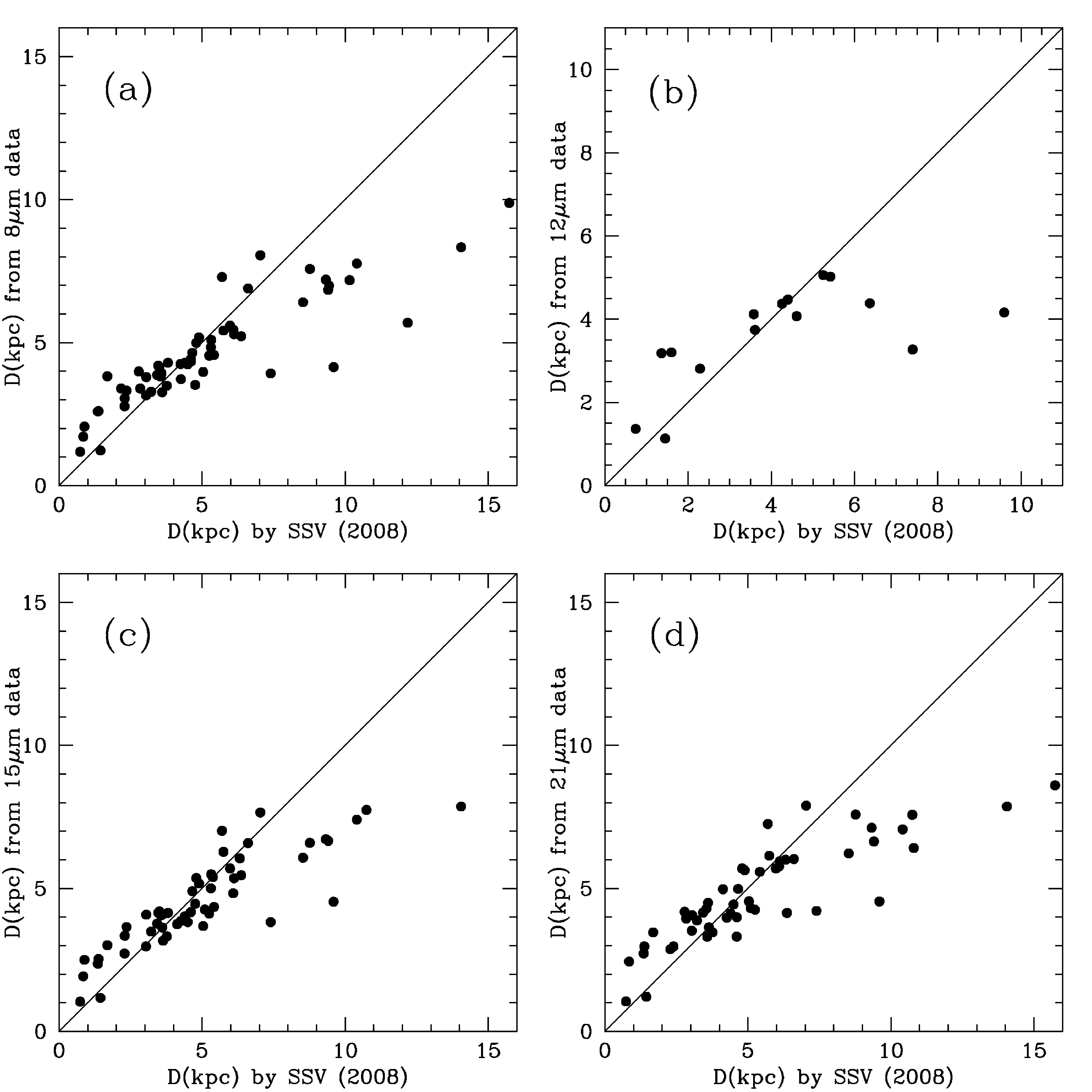}
\caption{Distances of PNe determined by the present method plotted
against the sample of calibrators compiled by SSV, for 4 the {\sc MSX}
bands: $8\,\mu$m (a), $12\,\mu$m (b), $15\,\mu$m (c), and $21\,\mu$m (d).}
\end{figure*}

\section{The calibration of a distance scale based on mid-infrared data}

Statistical distances are generally based on the hypothesis that
some nebular properties are constant or they can be evaluated
from distance-independent data. The Shklovsky (1956)
method for example, assumes that the ionized mass of the PN is
invariant from object to object, and as a result, the distance can
be expressed basically as a function of the incoming flux $F_{\lambda}$
and the nebular radius (Cahn \& Kaler 1971, Milne \& Aller 1975,
Milne 1982, Daub 1982).

Apart from secondary effects involving physical parameters of
the gas, such as electron temperature, the filling factor and the
abundance of helium, the distance of a PN can be expressed by the
following general equation:

\begin{equation}
D \propto F_{\lambda}^{\alpha} {\theta}^{\beta},
\end{equation}

\noindent where $F_{\lambda}$ is the flux density $-$ or the brightness
temperature $-$ $\theta$ is the nebular radius, and the constants
$\alpha$ and $\beta$ depend on the wavelength considered. The
empirical linear relationship between the radio and the infrared
intensities discussed in Sect. 2 allows us to use the {\sc MSX} flux densities
to calibrate a new distance scale based on the parameters above.
Equation (3) can be re-written in the following form:

\begin{equation}
\log D_{\rm MIR} = a_{\lambda} \log F_{\lambda} + b_{\lambda} \log {\theta}
+ c_{\lambda},
\end{equation}

\noindent where $D_{\rm MIR}$ is the distance of the nebula, $F_{\lambda}$
is the MIR flux density, $\theta$ is the angular radius, and the constants
a$_{\lambda}$, b$_{\lambda}$ and c$_{\lambda}$ depend on the wavelength.

\begin{table*}
\centering
\caption{Distances of {\sc MSX} counterparts of PNe listed in Ortiz
et al. (2005) but not listed as a calibrator by Stanghellini et al.
(2008, SSV), in alphabetical order.}
\begin{tabular}{clccrrrrrrrrr}
\hline
\#  & PN & {\sc MSX} & $\theta$ & F$_8$ & F$_{12}$ &
F$_{15}$ & F$_{21}$ & D$_{8}$ & D$_{12}$ & D$_{15}$ &
D$_{21}$ & D$_{\rm ave}$  \\
 & name & name & (\arcsec ) & (Jy) & (Jy) & (Jy) & (Jy) & (kpc) & (kpc) &
(kpc) & (kpc) & (kpc) \\

\hline

  1 & H1-16        & 000.1899+04.3725 &  0.90 &   0.16 &    -   &   2.18 &   6.09 &  8.93 &   -   &  7.12 &  7.28 &  7.78 \\
168 & H1-17        & 358.3515+03.0903 &  0.45 &   0.47 &   1.52 &   3.15 &  10.28 &  9.75 &  7.79 &  8.16 &  8.05 &  8.44 \\
165 & H1-18        & 357.6250+02.6032 &  0.75 &    -   &    -   &   1.76 &   3.95 &   -   &   -   &  7.98 &  8.56 &  8.27 \\
172 & H1-19        & 358.9684+03.3993 &  0.70 &   0.15 &   1.45 &   0.95 &   4.99 & 10.04 &  7.05 &  9.51 &  8.28 &  8.72 \\
171 & H1-20        & 358.9850+03.2281 &  1.65 &   0.16 &    -   &   1.65 &   3.26 &  7.03 &   -   &  6.25 &  6.94 &  6.74 \\
164 & H1-23        & 357.6075+01.7961 &  1.25 &   0.86 &    -   &   1.95 &   3.51 &  5.88 &   -   &  6.57 &  7.47 &  6.64 \\
151 & H1-31        & 355.1810$-$02.9147 &  0.35 &    -   &    -   &    -   &   3.07 &   -   &   -   &   -   & 11.64 & 11.64 \\
154 & H1-35        & 355.7329$-$03.4718 &  0.55 &   1.55 &   3.46 &   7.14 &  31.77 &  7.32 &  5.91 &  6.23 &  5.76 &  6.31 \\
148 & H1-36        & 353.5114$-$04.9194 &  0.40 &    -   &    -   &  14.81 &  24.70 &   -   &   -   &  5.77 &  6.78 &  6.28 \\
181 & H1-40        & 359.7140$-$02.6919 &  0.65 &   1.08 &   1.81 &   4.74 &  16.86 &  7.30 &  6.76 &  6.53 &  6.35 &  6.73 \\
 10 & H1-47        & 001.2950$-$03.0396 &  1.25 &    -   &    -   &    -   &   4.06 &   -   &   -   &   -   &  7.21 &  7.21 \\
  5 & H2-11        & 000.7117+04.7025 &  0.75 &   0.16 &    -   &    -   &   4.34 &  9.65 &   -   &   -   &  8.37 &  9.01 \\
 21 & H2-17        & 003.1935+03.4110 &  2.00 &    -   &    -   &    -   &   3.18 &   -   &   -   &   -   &  6.56 &  6.56 \\
 18 & H2-20        & 002.7963+01.6947 &  1.70 &   0.13 &    -   &    -   &   3.03 &  7.20 &   -   &   -   &  7.00 &  7.10 \\
 41 & Hb6          & 007.2758+01.8442 &  3.00 &   1.08 &    -   &   7.84 &  17.64 &  4.02 &   -   &  3.48 &  3.82 &  3.77 \\
  4 & He2-250      & 000.7249+03.2473 &  2.50 &   0.12 &    -   &   1.12 &    -   &  6.31 &   -   &  6.00 &   -   &  6.16 \\
 73 & K3-6         & 031.0357+04.1239 &  0.35 &   0.55 &   1.90 &   3.08 &  12.11 & 10.45 &  7.82 &  8.92 &  8.39 &  8.89 \\
 78 & M1-12        & 235.3839$-$03.9208 &  0.90 &   0.90 &   2.50 &   3.02 &   7.99 &  6.64 &  5.70 &  6.57 &  6.82 &  6.43 \\
173 & M1-26        & 358.9597$-$00.7209 &  1.60 &   5.27 &  15.88 &  29.26 & 170.40 &  3.90 &  2.97 &  3.09 &  2.73 &  3.17 \\
156 & M1-30        & 355.9023$-$04.2676 &  1.75 &   0.16 &    -   &   1.16 &   4.27 &  6.92 &   -   &  6.69 &  6.39 &  6.66 \\
 17 & M1-37        & 002.6807$-$03.4694 &  1.25 &   0.14 &    -   &    -   &   5.83 &  8.05 &   -   &   -   &  6.61 &  7.33 \\
 44 & M1-40        & 008.3356$-$01.1042 &  2.15 &   1.25 &   2.21 &   8.86 &  16.92 &  4.47 &  4.72 &  3.77 &  4.30 &  4.31 \\
 51 & M1-45        & 012.6066$-$02.7011 &  1.25 &   0.13 &    -   &   1.16 &   5.76 &  8.13 &   -   &  7.47 &  6.63 &  7.41 \\
142 & M2-4         & 349.8008+04.4649 &  1.00 &   0.19 &    -   &   1.99 &   4.15 &  8.37 &   -   &  7.04 &  7.71 &  7.70 \\
 25 & M2-14        & 003.6194+03.1836 &  1.10 &   0.29 &   1.40 &   1.42 &   6.48 &  7.49 &  6.34 &  7.42 &  6.72 &  6.99 \\
 37 & M2-31        & 006.0403$-$03.6209 &  2.00 &   0.19 &    -   &   1.73 &   3.86 &  6.39 &   -   &  5.80 &  6.26 &  6.15 \\
 58 & M3-25        & 019.7518+03.2739 &  0.65 &   0.75 &   1.78 &   5.35 &  15.11 &  7.77 &  6.79 &  6.34 &  6.52 &  6.85 \\
179 & M3-44        & 359.3842$-$01.8140 &  2.00 &    -   &    -   &   1.17 &  12.36 &   -   &   -   &  6.39 &  4.75 &  5.57 \\
 56 & M4-8         & 018.9038+03.6892 &  0.65 &   0.09 &    -   &    -   &   4.41 & 11.32 &   -   &   -   &  8.74 & 10.03 \\
 57 & M4-10        & 019.2348$-$02.2459 &  0.60 &   0.11 &    -   &   1.40 &   3.43 & 11.22 &   -   &  9.08 &  9.52 &  9.94 \\
 46 & NGC6537      & 010.0977+00.7396 &  2.35 &   3.64 &   4.50 &  18.12 &  40.34 &  3.58 &  3.80 &  3.06 &  3.40 &  3.46 \\
 68 & Pe1-18       & 027.3446$-$02.1124 &  0.60 &   0.42 &    -   &   2.20 &   8.78 &  8.88 &   -   &  8.12 &  7.61 &  8.20 \\
 16 & Th3-27       & 002.6923+04.2158 &  1.60 &   0.13 &    -   &    -   &    -   &  7.45 &   -   &   -   &   -   &  7.45 \\

\hline
\end{tabular}
\end{table*}

A comparison among the various statistical distances existing in the
literature shows that some of them have significant biases when
compared with each other. These effects can be minimized by
calibrating the distance scale with objects in the Galactic bulge
and/or in the Magellanic Clouds, which have distances determined
more accurately. The method proposed by Cahn et al. (1992) was
based on Shklovsky's (1956) and improved by Daub (1982). In that study, 36
Galactic PNe with distances accurately determined by several methods
was used to calibrate an empirical relationship between the distance
and the nebular flux (of H$\beta$ or at 5 GHz, when available) and
$\theta$. Eventually, that scale was re-calibrated using
the Magellanic Clouds as a distance standard (Stanghellini et al.
2008, hereafter SSV) and as a result they obtained a method
that describes the distance as a function of $T_b$ at 5 GHz and
$\theta$. SSV also produced a list of distances of Galactic PNe
to be used as calibrators in future distance scales.

The distance scale proposed in the present work was calibrated
with the catalogue of Galactic PNe distances by SSV.
Only PNe that appear both in Table 1 of SSV as well as in the
compilation of {\sc MSX} counterparts of PNe (Table 3 of Ortiz
et al. 2005) have been considered (67 objects, Table 2). PNe with
bad-quality {\sc MSX} photometry and/or without size measurements
have been discarded. In order to obtain the coefficients of equation
4, we fitted the SSV distances, {\sc MSX} flux densities and
the nebular radii using the least-squared method. The nebular radii
$\theta$ used in this work have been preferentially extracted
from Si\'odmiak \& Tylenda (2001), obtained with the VLA.
In the cases where this measurement was not available, $\theta$
was taken from the compilation of SSV. Among the 67 PNe used
in the calibration, there are only 26 objects with apparent
radii listed both in Siodmiak \& Tylenda as well as in SSV and,
among these, there are only 4 discrepant determinations of $\theta$:
M2-16, M3-14, Hb5, and NGC2440. Apart from these few exceptions the
agreement between the two radii (optical and radio) is excellent.

Table 1 presents the MSX bands, their isophotal wavelength
(Egan et al. 1999) and the fitting coefficients obtained with the
least-squared method, where the distance $D$ must be given in
kiloparsecs, $F_{\lambda}$ in Jansky, and $\theta$ in arc seconds.
Because the flux quality varies from object to object and from band
to band, the number $N$ of PNe used in each fit is not the same.
A simple examination of Table 1 reveals
that the coefficients $a_{\lambda}$, $b_{\lambda}$ and $c_{\lambda}$
show a non-negligible variation with the wavelength. The values
relative to 12 $\mu$m are the most uncertain because of the
small number of objects with good-quality photometry, caused by
the lack of sensitivity in this band. If one takes the average value
of these coefficients, equation 3 would be written with
$\langle\alpha\rangle=-0.23$ and $\langle\beta\rangle=-0.32$.

\begin{figure*}
\includegraphics[width=16.0cm]{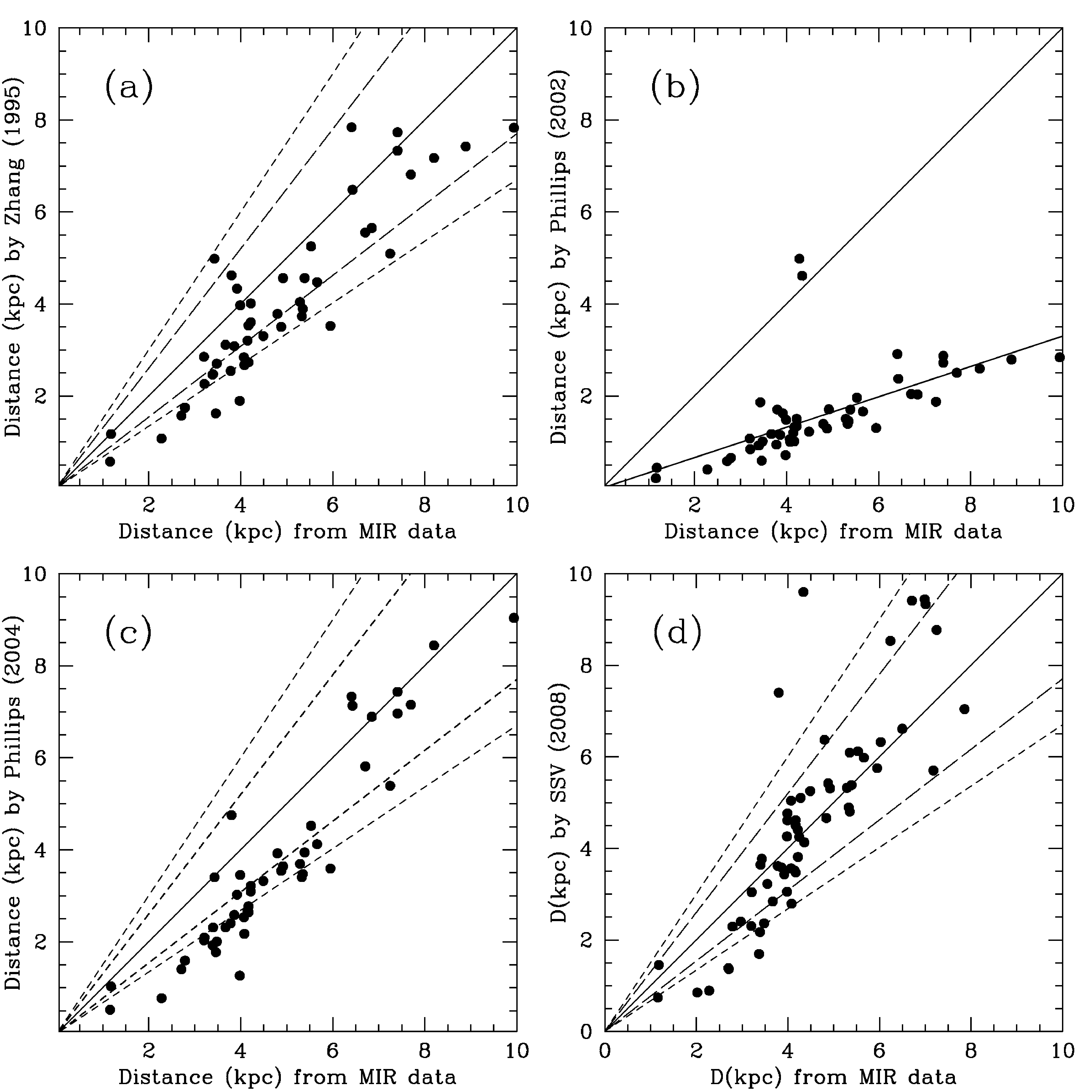}
\caption{Distances of PNe determined by the present method plotted
against other distance scales by: (a) Zhang (1995), (b) Phillips
(2002), (c) Phillips (2004), and (d) Stanghellini et al. (2008, SSV).
The auxiliary dashed lines drawn on both sides of the identity line
represent a deviation of 30\% and 50\% from equality.}
\end{figure*}

\section{Results}

In Table 2, we list the original data used in the fitting of equation 4
and the various distances, obtained by SSV and in this work.
The columns are: (1) Ordering number,
according to Ortiz et al. (2005); (2) Name; (3) {\sc MSX} name; (4) Angular
radius (\arcsec ); (5$-$8) {\sc MSX} flux densities (Jansky); (9-12) Distance
(in kpc) obtained from the {\sc MSX} bands centred at 8, 12, 15 and 21 $\mu$m,
respectively; (13) Average Distance (kpc); (14) Distance according to SSV.
A comparison between $D_\mathrm{SSV}$ and $D_\mathrm{ave}$
is shown in Fig. 3d. The points are approximately dispersed near the
identity line within $\pm$50\%. The scattering of the points results
mainly from two sources of errors: {\it (i)} intrinsic errors of the
present method (including errors in the data);
{\it (ii)} errors in $D_{\rm SSV}$, estimated as
$\sim$30\% in the original reference. According to the authors,
this figure is similar to that observed in other distance scales
and cannot be significantly reduced by using better calibrators.

Among the factors that contribute to the scattering seen in
Fig. 3 is the error associated with the nebular radius $\theta$.
Fortunately, the majority of the
$\theta$ measurements used in our fit were determined with the VLA,
which gives spatial resolution superior to any optical measurement
obtained by ground-based telescopes. The error in the determination
of $D_\mathrm{MIR}$ caused by an error in $\theta$ is:

\begin{equation}
\left( \frac{{\sigma}_D}{D}\right)_\mathrm{MIR} = \vert b_{\lambda}\vert
\frac{{\sigma}_{\theta}}{\theta},
\end{equation}

\noindent where ${\sigma}_{\theta}$ is the error bar associated
with the measurement of the nebular radius. Assuming ${\sigma}_{\theta}=
1$ \arcsec -- corresponding to a typical seeing at the visual 
wavelengths -- the error in  $D_\mathrm{MIR}$ calculated with equation
5 and $b_{\lambda}$ taken from Table 1 is 30\% -- 40\% for a PN with
$\theta = 1$\arcsec. This error drops down to a half of this value
for a PN with $\theta = 2$\arcsec and so on.

The errors in the {\sc MSX} flux densities are small, between 3\%
and 6\% for the bands listed in Table 1 (Egan et al. 1999). Thus, this
source of error does not play an important role in the evaluation of
the total error in $D$, especially when compared to the contribution
of the error in $\theta$.

\begin{table*}
\centering
\caption{Distances of PNe observed by {\sc GLIMPSE}. The original data were
extracted from Kwok et al. (2008, first set of data) and Zhang \& Kwok
(2009, second set below). The columns are (1) {\sc PNG} name; (2) other name,
(3) {\sc MSX} name; (4) integrated flux density (Jansky) at 8 $\mu$m, corrected
(see text); (5) nebular radius, in arc seconds; (6) $D_{\rm MIR}$ (kpc).}

\begin{tabular}{clcrrc}
\hline
{\sc PNG} name & Other name & {\sc MSX} name & F$_{8}$(Jy) & $\theta$(\arcsec )
 & $D_{\rm MIR}$(kpc) \\
\hline

PNG010.1+00.7 & NGC6537     & MSX010.0997+00.7396 &   4.968 &  6.0 &  2.35 \\
PNG011.7$-$00.6 & NGC6567     & MSX011.7438$-$00.6502 &   2.225 &  6.0 &  2.71 \\
PNG018.6$-$00.0 & G018.6$-$00.0 &  &  40.508 & 60.0 &  0.67 \\
PNG035.5$-$00.4 & G035.5$-$00.4 & MSX035.5653$-$00.4913 &   0.155 &  5.0 &  4.61 \\
PNG040.3$-$00.4 & G040.3$-$00.4 &  &   0.198 & 10.0 &  3.37 \\
PNG055.5$-$00.5 & M1$-$71       & MSX055.5067$-$00.5579 &   1.290 &  5.0 &  3.19 \\
PNG056.4$-$00.9 & K3$-$42       &  &   0.052 &  3.5 &  6.42 \\
PNG062.4$-$00.2 & M2$-$48       &  &   0.028 &  3.5 &  7.16 \\
PNG295.7$-$00.2 & G295.7$-$00.2 &  &   8.294 & 30.0 &  1.15 \\
PNG296.8$-$00.9 & G296.8$-$00.9 &  &   0.029 &  5.0 &  6.19 \\
PNG298.4+00.6 & G298.4+00.6 &  &  23.292 & 90.0 &  0.63 \\
PNG300.2+00.6 & He2$-$83      & MSX300.2787+00.6627 &   0.331 &  5.0 &  4.04 \\
PNG300.4$-$00.9 & He2$-$84      &  &   0.025 &  5.0 &  6.32 \\
PNG315.0$-$00.3 & He2$-$111     & MSX315.0301$-$00.3701 &   0.267 &  7.5 &  3.58 \\
PNG318.9+00.6 & G318.9+00.6 & MSX318.9321+00.6956 &   0.166 & 10.0 &  3.48 \\
PNG333.9+00.6 & G333.9+00.6 & MSX333.9294+00.6858 &   1.057 & 22.5 &  1.84 \\
PNG343.9+00.8 & H1$-$5        & MSX343.9919+00.8347 &   0.869 &  6.0 &  3.19 \\
              &             &                     &         &      &       \\
PNG000.0$-$01.3 & G000.0$-$01.3 & &   0.022 &  5.8 &  6.11 \\
PNG000.1$-$01.7 & G000.1$-$01.7 & &   0.011 &  6.0 &  6.80 \\
PNG000.3$-$01.6 & G000.3$-$01.6 & &   0.011 &  4.0 &  7.97 \\
PNG000.5+01.9 & K6$-$7        & &   0.029 &  5.0 &  6.19 \\
PNG000.9$-$01.0 & G000.9$-$01.0 & &   0.025 &  4.5 &  6.59 \\
PNG000.9$-$01.8 & G000.9+01.8 & &   0.013 &  2.5 &  9.27 \\
PNG001.0$-$01.9 & K6$-$35       & &   0.019 &  6.5 &  6.01 \\
PNG001.6$-$01.1 & G001.6$-$01.1 & &   0.033 &  4.8 &  6.13 \\
PNG002.0+00.7 & G002.0+00.7 & &   0.043 &  4.0 &  6.29 \\
PNG002.1$-$00.9 & K5$-$35       & MSX002.1198$-$00.9610 &   0.238 &  6.0 &  3.99 \\
PNG002.1$-$01.1 & G002.1$-$01.1 & &   0.016 &  3.0 &  8.31 \\
PNG002.2$-$01.2 & G002.2$-$01.2 & &   0.024 &  5.5 &  6.14 \\
PNG003.4+01.4 & G003.4+01.4 & &   0.021 &  5.0 &  6.53 \\
PNG003.5$-$01.2 & G003.5$-$01.2 & MSX003.5794$-$01.2219 &   1.188 &  5.5 &  3.12 \\
PNG003.5+01.3 & G003.5+01.3 & MSX003.5562+01.3532 &   0.057 &  5.0 &  5.49 \\
PNG003.6$-$01.3 & G003.6$-$01.3 & &   0.025 &  5.0 &  6.32 \\
PNG004.3$-$01.4 & G004.3$-$01.4 & &   0.032 &  4.0 &  6.62 \\
PNG004.8$-$01.1 & G004.8$-$01.1 & MSX004.8343$-$01.1940 &   0.133 &  5.8 &  4.47 \\
PNG006.1+00.8 & G006.1+00.8 &  &   0.030 &  5.5 &  5.92 \\
PNG353.9+00.0 & G353.9+00.0 &  &   0.065 &  5.0 &  5.37 \\
PNG355.6+01.4 & G355.6+01.4 & MSX355.6143+01.4111 &   0.095 &  5.0 &  5.03 \\
PNG356.0$-$01.4 & G356.0$-$01.4 &  &   0.032 &  5.5 &  5.85 \\
PNG356.9+00.9 & G356.9+00.9 & MSX356.9505+00.9112 &   0.228 &  5.0 &  4.32 \\
PNG357.5+01.3 & G357.5+01.3 &  &   0.105 &  5.0 &  4.94 \\
PNG357.7+01.4 & G357.7+01.4 &  &   0.012 &  3.0 &  8.77 \\
PNG358.2$-$01.1 & Al2$-$L       &  &   0.062 &  9.5 &  4.22 \\
PNG358.8$-$00.0 & Terz2022    & &  55.728 & 40.0 &  0.74 \\
PNG359.1$-$01.7 & He1$-$191     & MSX359.1155$-$01.7195 &   0.303 &  7.5 &  3.51 \\

\hline
\end{tabular}
\end{table*}

Table 3 contains the distances calculated for PNe present in the
sample of Ortiz et al. (2005), but absent in the list of SSV.
The columns are: (1) Ordering number,
according to Ortiz et al. (2005); (2) Name; (3) {\sc MSX} name; (4) Angular
radius (\arcsec ); (5$-$8) {\sc MSX} flux densities (Jansky); (9-12) Distance
(in kpc) obtained from the flux density at 8, 12, 15 and 21 $\mu$m,
respectively; (13) Average Distance (kpc).
In order to compare our results with others, we merged Tables
2 and 3 and plotted the average distance obtained by the present
method against other values given by methods based on radio fluxes,
extracted from several references. The distance scale proposed by
Zhang (1995) was based on a correlation between $T_b$ and
the nebular radius (an upgrade to the Shklovsky method), and
calibrated with a sample of 134 ``standard'' nebulae.
Figure 3 shows that Zhang's distance scale is about $\sim 20$\%
shorter than ours (and SSV, by extension). Phillips (2002) proposed
a statistical method similar to Zhang's (1995), but calibrated it
with a sample of nearby ($D < 0.7$ kpc) PNe.
His distance scale was based on a revised relationship between the
surface brightness and the nebular radius and, when compared to our
results (Fig. 3, top-right), his method produces distances shorter by a
factor $\sim 3$. A revision of that method was eventually proposed by Phillips
(2004), that assumes a relationship between the nebular luminosity at 5
GHz and the nebular brightness temperature, $T_b$. A comparison
between our results and the distances produced by the latter method
($D_{\rm Ph04}$) is presented in the bottom-left panel of Fig. 3:
there seems to be a bias associated with $D_{\rm MIR} < 6$ kpc, but
both methods agree within $\sim 30$\% beyond 6 kpc.

\subsection{Distances obtained from other mid-infrared surveys:
{\it Spitzer}}

The present method was calibrated with nebular flux densities
of the {\sc MSX} survey centred at 8.3, 12.1, 14.7 and
21.3 $\mu$m.  This calibration (equation 4 and Table 1)
can be extended to other surveys
if their effective wavelength is similar to {\sc MSX}. As an example, we
present in this section distances of PNe detected by the {\it Spitzer}
satellite during the {\it GLIMPSE Legacy Survey} (Fazio et al. 2004).
As mentioned in the introduction, the {\it IRAC} camera had one band
centred at $\lambda = 8.0\,\mu$m, not very different from the {\sc MSX} band 'A'
centred at $\lambda = 8.3\,\mu$m. Differently from ``point-source surveys'',
{\sc GLIMPSE} recorded full images of PNe and the flux density integrated
over the total nebular solid angle can be easily computed and used to
produce distances by our method.

Kwok et al. (2008) and Zhang \& Kwok (2009) obtained integrated flux
densities of over 60 PNe with sensitivity superior to {\sc MSX}. A
comparison between the {\sc GLIMPSE} and {\sc MSX} fluxes of a sample of 22
PNe showed the existence of a bias expressed as:
$F({\rm IRAC\,8.0})/F({\rm MSX\,8.3})=0.90 \pm 0.36$ (Zhang \& Kwok 2009).
We applied this correction
factor to {\sc GLIMPSE} flux densities at 8.0 $\mu m$ to calculate $D_{\rm MIR}$
and listed these results in Table 4. The first set of data, from row
1 to 17, refers to PNe situated along the Galactic disc, between
$10^\circ - 60^\circ$ away from the Galactic centre. The second set, from the
18th row on, refers to PNe situated
within $5^\circ$ from the Galactic centre, i.e. seen against the bulge.
A simple inspection of these two sets of objects shows that, on average
``disc'' PNe have smaller distances when compared to ``bulge'' objects.
Actually the distances of the ``disc'' PNe show a wide range of values,
from 0.6 to 7.2 kpc. On the other hand, the distances of the ``bulge''
PNe are more homogeneous, from 3.1 to 9.3 kpc. The only exception is
the object named PNG358.8-00.0 (Terz2022): considering its large
infrared flux and apparent size, it is a clear example of a nearby object
seen against the bulge. Table 4 also shows how deep is the {\it Spitzer}
survey when compared to {\sc MSX}: the faintest flux density at 8 $\mu$m
in Tables 2 and 3 is $F_8=0.11$ Jansky (object M4-10), whereas Table 4
contains plenty of nebulae with one-tenth of this value.

\section{Conclusions}

In this work we proposed a distance scale of planetary nebulae based
on mid-infrared flux densities and the nebular apparent size. A study
of two distance-independent nebular parameters -- the brightness
temperature at 5 GHz and the mid-infrared specific intensity $I_{\lambda}$ --
showed that these two characteristics are very well correlated.
This relationship, as well as the similar formulation of the
distance scale -- $D$ as a function of the nebular monochromatic
specific intensity
and its size -- allows us to state that the reliability of our
method is similar to SSV's. In fact, a comparison between the nebular
distances calculated by both methods showed that they produce
similar results within 50\%. Nevertheless, since both methods are
statistical, each of them might be accurate only within the natural
limitation imposed by the set of hypotheses assumed to be
valid for the whole set of PNe, but that may vary significantly
from object to object. These variations include the different effect
of the various spectral features (continuum, atomic lines, molecular bands)
on the final flux, deviations from sphericity, etc. However, like
in many statistical methods, particular deviations from the average
may not introduce significant errors in the distance scale,
but only in the error bars. As an example, according
to SSV, excluding bipolar PNe from the sample would not change
their distance scale by more than 5\%, and a similar figure is
expected here. On the other hand, variations of this kind may affect
the distance error bars, and a comparison between the distances
determined by SSV and our results showed that are equal within
$\pm$50\%.

Although the effects of the [C/O] ratio -- either
of the gas or on the grain surface -- on the nebular spectrum are well known,
we could not detect its effect on the distance scale proposed, because
of the small number of objects in our sample with observed C and/or O
molecular features.

As an example of application of the present method, we derived distances
of a sample of PNe observed by the {\it Spitzer} infrared telescope.
This distance scale can be extended to any other database containing
nebular radii and flux densities near 8, 12, 15, and 21 $\mu$m, such
as {\sc IRAS}, {\sc ISOGAL}, and future infrared surveys of this kind.

\section*{Acknowledgements}

We thank Professor W.J. Maciel for his collaboration in the early phase
of this work. This work has been partially funded by the S\~ao Paulo
Agency for Science Support {\sc FAPESP}, grant no. 2010/18835-3.
This research has made use of NASA's Astrophysics Data System
and the {\sc SIMBAD} Astronomical Database.

\end{document}